# Radiative corrections to anti-neutrino proton scattering

**Udit Raha,**[a] **Fred Myhrer**[*b] **and Kuniharu Kubodera**[b]

[a]*Indian Institute of Technology Guwahati, 781 039 Assam, India*
[b]*Dept. Physics and Astronomy, University of South Carolina, Columbia, SC 29208, USA.*
*E-mail:* Udit.Raha@unibas.ch, myhrer@sc.edu, kubodera@physics.sc.edu

The inverse $\beta$-decay reaction, $\bar{\nu}_e p \to e^+ n$, for low-energy anti-neutrinos coming from nuclear reactors is of great current interest in connection with high-precision measurements of the neutrino mixing angle $\theta_{13}$. We have derived analytic expressions, up to next-to-leading order in heavy-baryon chiral perturbation theory, for the radiative corrections (RCs) and the nucleon-recoil corrections both for this reaction and for the related neutron $\beta$-decay process. We show that the recoil corrections, which include the "weak magnetism" contribution, are small for neutron $\beta$-decay, but for inverse $\beta$-decay, the recoil corrections are comparable in size to the RCs for typical energies of reactor anti-neutrinos, and they have opposite signs. The RCs and the recoil corrections exhibit very different dependences on the neutrino energy.

*Speaker.

Very recently, several experimental collaborations reported nonzero values of the neutrino mixing parameter, $\theta_{13}$ [1, 2, 3, 4]. Low-energy anti-neutrinos from nuclear reactors are well suited to determining $\theta_{13}$. The Double-Chooz [1], Daya Bay [2], and RENO [3] Collaborations have been carrying out high-precision measurements of $\theta_{13}$, by monitoring the inverse $\beta$-decay reaction $\bar{\nu}_e + p \to e^+ + n$ for the reactor-generated $\bar{\nu}_e$'s. The accurate extraction of $\theta_{13}$ from a measured positron yield in this reaction requires a precise knowledge of radiative corrections (RCs) and nucleon-recoil corrections. An important issue in this connection is to what extent one can exploit the experimental data on neutron $\beta$-decay $n \to p + e^- + \bar{\nu}_e$ to have a better control of the RCs and recoil corrections to the inverse $\beta$-decay reaction.

In earlier works [5, 6, 7], the relevant RCs were evaluated to order-$\alpha$ in the theoretical framework developed by Sirlin and Marciano, to be referred to as the S-M approach; see, *e.g.*, Refs. [8, 9]. Although the estimates based on the S-M approach are believed to be reliable to the level of accuracy quoted in the literature, it is not totally excluded that these estimates may involve some degree of model dependence. Meanwhile, the nucleon-recoil corrections for the inverse $\beta$-decay process were evaluated based on the $1/m_N$-expansion of the nucleon weak-interaction form factors, where $m_N$ is the nucleon mass; see e.g., Refs. [6, 10].

In Refs. [11, 12], we proposed to use an effective field theory (EFT) approach to derive the RCs and nucleon-recoil corrections for the two processes. Analytic expressions for both the RCs and the recoil corrections to next-to-leading order (NLO) were presented for neutron $\beta$-decay [11], and for inverse $\beta$-decay [12]. Our analytic RC expressions are consistent with those obtained in the S-M approach by Fukugita and Kubota [5], and by Sirlin [13]; see also Refs. [6, 7] for earlier works. As for the recoil corrections, our analytic results for inverse $\beta$-decay presented in Ref. [12] were found to be consistent with Ref. [10]. We remark, however, that the recoil corrections (in particular, the one arising from the weak-magnetism term) have very different analytic expressions for neutron $\beta$-decay and the inverse $\beta$-decay reaction. In this note we summarize some analytic results of Refs. [11, 12], and present some numerical examinations which we hope will shed further light on the practical significance of the RCs and the recoil corrections for neutron $\beta$-decay and inverse $\beta$-decay.

EFT allows a unified approach to electro-weak and strong processes at low energies in a gauge and model independent way. Let $\bar{Q}$ represent the typical four-momentum involved in neutron $\beta$-decay, or inverse $\beta$-decay, where $\bar{Q}$ is very small even compared with the pion mass, so long the incident anti-neutrinos are those coming from nuclear reactors. In particular, the well-established chiral perturbation theory ($\chi$PT), see e.g., Refs. [14, 15, 16], provides an ideal framework for a study of the two processes. This EFT framework admits a perturbative expansion of relevant Feynman amplitudes in terms of two expansion parameters: (i) the $\chi$PT expansion parameter, $\bar{Q}/\Lambda_\chi$, where $\Lambda_\chi \simeq 4\pi f_\pi \approx 1$ GeV is the chiral scale; (ii) the usual QED expansion parameter, $\alpha/(2\pi)$, which governs the contributions from the electroweak RCs. We adopt here heavy-baryon chiral perturbartion theory (HB$\chi$PT), e.g., Ref. [14]. In this scheme an additional expansion parameter, $\bar{Q}/m_N$, governs the contributions from nucleon-recoil corrections. Since $\Lambda_\chi \simeq m_N \approx 1$ GeV, the $\bar{Q}/\Lambda_\chi$ and $\bar{Q}/m_N$ can be considered as one expansion parameter, and thus, is a natural part of our EFT. In EFT the short-distance physics, which is not probed in low-energy processes, is subsumed into so-called *low-energy constants* (LECs). Although in principle the LECs can be evaluated from the fundamental theory, i.e., QCD, a more pragmatic approach is to determine the LECs by fitting experimental observables. Once the LECs are known, we can make predictions for other measurable quantities. Our concern here is to carry out a HB$\chi$PT-based calculation of the RCs and the recoil corrections up to NLO, *viz.*, first order in $\alpha/2\pi$, $\bar{Q}/\Lambda_\chi$, and $\bar{Q}/m_N$.

The effective lagrangian to NLO relevant to our calculation is given by

$$\mathcal{L}_{eff} = \mathcal{L}_{QED} + \mathcal{L}_{NN} + \mathcal{L}_{NN\psi\psi}, \tag{1}$$

$$\mathcal{L}_{QED} = -\frac{1}{4}F^{\mu\nu}F_{\mu\nu} - \frac{1}{2\xi_A}(\partial\cdot A)^2 + \left(1+\frac{\alpha}{4\pi}e_1\right)\bar{\psi}_e(i\gamma\cdot D)\psi_e + m_e\bar{\psi}_e\psi_e + \bar{\psi}_\nu i\gamma\cdot\partial\psi_\nu, \tag{2}$$

$$\mathcal{L}_{NN} = \bar{N}\left[1+\frac{\alpha}{8\pi}e_2(1+\tau_3)\right](iv\cdot D)N, \tag{3}$$

$$\mathcal{L}_{NN\psi\psi} = -\left(\frac{G_F V_{ud}}{\sqrt{2}}\right)[\bar{\psi}_e\gamma_\mu(1-\gamma_5)\psi_\nu]\left\{\bar{N}\tau^+\left[\left(1+\frac{\alpha}{4\pi}e_V\right)v^\mu - 2g_A\left(1+\frac{\alpha}{4\pi}e_A\right)S^\mu\right]N\right.$$
$$\left. +\frac{1}{2m_N}\bar{N}\tau^+\left[i(v^\mu v^\nu - g^{\mu\nu})(\overleftarrow{\partial}-\overrightarrow{\partial})_\nu - 2i\mu_V\left[S^\mu, S\cdot(\overleftarrow{\partial}+\overrightarrow{\partial})\right] - 2ig_A v^\mu S\cdot(\overleftarrow{\partial}-\overrightarrow{\partial})\right]N\right\}. \tag{4}$$

The $\mathcal{L}_{QED}$ in Eq.(2) is the QED lagrangian, where $F_{\mu\nu} = \partial_\mu A_\nu - \partial_\nu A_\mu$, and $D_\mu = \partial_\mu + ieA_\mu$. The $\mathcal{L}_{NN}$ is part of the HB$\chi$PT lagrangian and includes photon-nucleon interactions, while $\mathcal{L}_{NN\psi\psi}$ represents low-energy LO and NLO weak interactions including the explicit forms of NLO nucleon-recoil terms dictated by HB$\chi$PT. Here $g_A$ is the axial coupling constant while $v_\mu$ is the nucleon four-velocity and $S^\mu$ is the nucleon spin. In the NLO part of the lagrangian, $\mu_V = \mu_p - \mu_n$ is the nucleon isovector magnetic moment. The low-energy constants (LECs), $e_1, e_2, e_V$ and $e_A$, incorporate the short-range radiative physics that is not probed in a low-energy process. The Fermi coupling constant, $G_F$ is determined from muon decay, and $|V_{ud}|$ is the CKM matrix element.

The EFT-based calculations of the NLO radiative and recoil corrections are described in Refs.[11] and [12] for neutron $\beta$-decay and inverse $\beta$-decay, respectively. We focus on the differential decay rate for neutron $\beta$-decay, $n(p_n) \to p(p_p) + e^-(p) + \bar{\nu}_e(p_\nu)$, and the differential cross section for the inverse $\beta$-decay reation, $\bar{\nu}_e(p_\nu) + p(p_p) \to e^+(p) + n(p_n)$, and assume an experimental situations where none of the particle spins are detected. Here the four-momenta of particles are indicated in the parentheses; for the four-momenta, $p$ and $p_\nu$, we shall also use quantities defined by $p = (E, \mathbf{p})$ and $p_\nu = (E_\nu, \mathbf{p}_\nu)$. The neutron $\beta$-decay differential decay rate is given in Ref. [11] and the differential cross-section for inverse $\beta$-decay is given in Ref. [12] as:

$$\frac{d\sigma_{inv\beta}}{d(\cos\theta_e)} = \left(\frac{G_F V_{ud}}{\sqrt{2}}\right)^2 f(E)\left[(1+3g_A^2)\mathcal{G}_1(\beta) + (1-g_A^2)\mathcal{G}_2(\beta)\beta\cos\theta_e\right], \tag{5}$$

where $\mathcal{G}_i(\beta)$'s ($i=1,2$) contain radiative and recoil corrections, $\beta = |\mathbf{p}|/E = \sqrt{E^2-m_e^2}/E$ the velocity of outgoing electron/positron, $\cos(\theta_e) = \hat{p}_\nu \cdot \hat{p}$, $f(E)$ the phase-space factor derived in [12][1]

$$f(E) = \frac{E^2\beta}{\pi}\left[1 - \frac{1}{m_N}\left(E - \frac{E_\nu}{\beta}\cos\theta_e\right) + \mathcal{O}(m_N^{-2})\right]. \tag{6}$$

The velocity-dependent functions, $\mathcal{G}_i(\beta)$ ($i = 1, 2$), contain RCs and nucleon-recoil corrections,

$$\mathcal{G}_i(\beta) = 1 + \frac{\alpha}{2\pi}\mathcal{G}_i^{rad}(\beta) + \frac{1}{m_N}\mathcal{G}_i^{recoil}(\beta). \tag{7}$$

---

[1] Note that for inverse $\beta$-decay, the positron energy, $E$, and velocity, $\beta$, contain terms of $\mathcal{O}(m_N^{-1})$, e.g., $E = \tilde{E}(1 - \mathcal{O}(m_N^{-1}))$, where $\tilde{E} = E_\nu - (m_n - m_p)$; see the discussion leading to Eq. (14) in Ref. [12].

The kinematical recoil ($m_N^{-1}$) corrections are included in the phase-space factor $f(E)$, and the recoil corrections in Eq.(7) are dynamical ones coming from Eq.(4). The HB$\chi$PT calculation of RCs was derived in [11] for neutron $\beta$-decay (denoted $\mathscr{K}_1^{rad}(\beta)$ below), and for inverse $\beta$-decay in [12].[2]

$$1 + \frac{\alpha}{2\pi}\mathscr{K}_1^{rad}(\beta) = \left[1 + \frac{\alpha}{2\pi}\tilde{e}_V^R(\mu^2)\right]\left[1 + \frac{\alpha}{2\pi}\left(\delta_\alpha^{(1)}(\beta) - \frac{5}{4}\right)\right] \quad (8)$$

$$1 + \frac{\alpha}{2\pi}\mathscr{G}_1^{rad}(\beta) = \left[1 + \frac{\alpha}{2\pi}\tilde{e}_V^R(\mu^2)\right]\left[1 + \frac{\alpha}{2\pi}\delta_{out}(\beta)\right] \quad (9)$$

The "inner" RCs, which are independent of $\beta$, are encoded in the LEC, $\tilde{e}_V^R(\mu^2)$. The "outer" RCs, $\delta_\alpha^{(1)}(\beta)$, and $\delta_{out}(\beta)$, are well-known, model-independent, long-distance QED corrections that contain no hadronic effects. Expressions for the outer corrections are in, e.g., Refs. [11, 12]. The ultraviolet-divergent and scale dependent terms are subsumed in LECs, and all infrared-divergent terms of $\mathscr{O}(\alpha)$ are simultaneously canceled, leading to finite final results. The recoil corrections, $\mathscr{K}_1^{recoil}(\beta)$ pertaining to neutron $\beta$-decay, and $\mathscr{G}_1^{recoil}(\beta)$ appearing in Eq.(7), are given in [11, 12]:

$$\mathscr{K}_1^{recoil}(\beta) = \beta^2 E\left(\frac{1 + 2g_A\mu_V + g_A^2}{1 + 3g_A^2}\right) + E_\nu\left(\frac{1 - 2g_A\mu_V + g_A^2}{1 + 3g_A^2}\right) + \mathscr{O}(m_N^{-1}) \quad (10)$$

$$\mathscr{G}_1^{recoil}(\beta) = \beta^2 E\left(\frac{1 - 2g_A\mu_V + g_A^2}{1 + 3g_A^2}\right) - E_\nu\left(\frac{1 + 2g_A\mu_V + g_A^2}{1 + 3g_A^2}\right) + \mathscr{O}(m_N^{-1}) \quad (11)$$

A noteworthy point in the above expressions is that terms involving $2g_A\mu_V \simeq 12$ arising from the "weak-magnetism" interactions in Eq.(4) constitute the most dominant recoil corrections.

Since these two processes involve the same LEC, $\tilde{e}_V^R(\mu^2)$, as the only unknown parameter, it is in principle possible to use neutron $\beta$-decay data to fix $\tilde{e}_V^R(\mu^2)$, and make model-independent predictions for the inverse $\beta$-decay cross sections. This statement is true to the extent that the other quantities which go into the above formulae and which are usually treated as "known" quantities are indeed known with high enough precision. To what degree the existing uncertainties in the values of $g_A$ and the neutron lifetime affect the outcome of such an analysis is a subject that deserves a careful study. Relegating this investigation to a future study, we choose here to use an estimate for $\tilde{e}_V^R(\mu^2)$ deduced in Ref. [11] by comparing the results for neutron $\beta$-decay obtained in HB$\chi$PT with those obtained in the S-M approach [8, 9]. This comparison yields $\tilde{e}_V^R(\mu^2 = m_N^2) \approx 4\ln\left(\frac{m_Z}{m_p}\right) + \ln\left(\frac{m_p}{m_A}\right) + 2C + A_g = 19.5 \pm 0.7$ where we allow a "liberal" choice of the $A_1$-resonance mass, $m_A \approx 1.2 \pm 0.6$ MeV, e.g., Ref. [17]. Thus the RC from LEC in Eqs. (8)$\sim$(9) is estimated to be $(\alpha/2\pi)\tilde{e}_V^R(m_N^2) \simeq 0.023$. This short-distance (inner) contribution to $\tilde{e}_V^R(m_N^2)$ is dominated by well-known electro-weak box diagrams [8, 9]. We use $\tilde{e}_V^R(\mu^2 = m_N^2) = 19.5 \pm 0.7$ in the following.

To illustrate interplay between the RCs and recoil corrections we concentrate on the angle-integrated observables, *viz.*, the differential decay rate, $d\Gamma_\beta/dE$, for $\beta$-decay, and the total cross section, $\sigma_{inv\beta}$, for inverse $\beta$-decay, i.e., we compare $\mathscr{K}_1^{rad}$, $\mathscr{G}_1^{rad}$, $\mathscr{K}_1^{recoil}$, and $\mathscr{G}_1^{recoil}$. In Fig.1 we plot $\frac{\alpha}{2\pi}\mathscr{K}_1^{rad}$ for neutron $\beta$-decay as a function of $E_\nu$ ($0 \leq E_\nu \leq 0.78$ MeV). The pair of RC curves in the figure indicates an error-band associated with the above-mentioned uncertainty in $\tilde{e}_V^R(\mu^2 = m_N^2)$. Fig.2 shows $\frac{\alpha}{2\pi}\mathscr{G}_1^{rad}$ for inverse $\beta$-decay versus $E_\nu$.

Fig.1 also shows $m_N^{-1}\mathscr{K}_1^{recoil}$ for $\beta$-decay, while Fig.2 includes $m_N^{-1}\mathscr{G}_1^{recoil}$ for the inverse $\beta$-decay. It is seen that $\mathscr{K}_1^{recoil}$ is much smaller than $\mathscr{G}_1^{recoil}$. The difference between $\mathscr{K}_1^{recoil}$ and

---

[2]It is to be remarked that the HB$\chi$PT results are consistent with those obtained in the the S-M approach [5, 6, 7, 8].

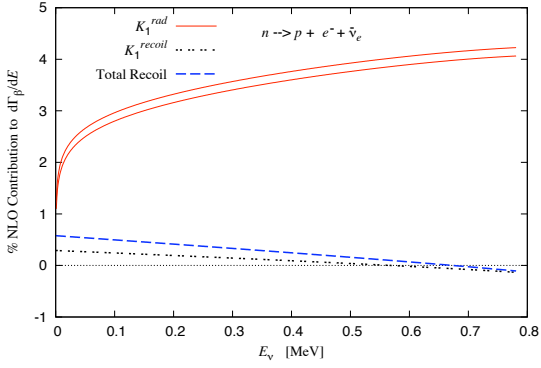
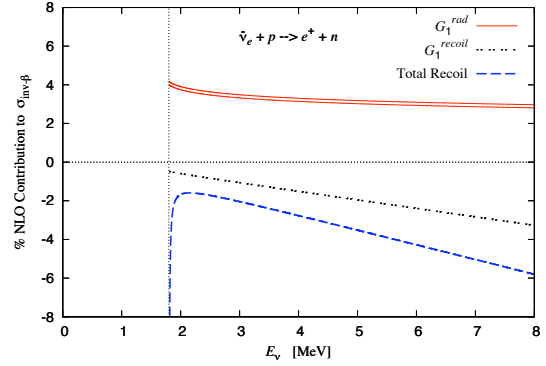

**Figure 1:** Radiative correction, $\frac{\alpha}{2\pi}\mathcal{K}_1^{rad}$, Eq. (8), and recoil correction, $m_N^{-1}\mathcal{K}_1^{recoil}$, Eq. (10), are plotted as functions of the emitted anti-neutrino energy $E_\nu$ for neutron $\beta$-decay. The pair of RC curves gives an "error band" that reflects uncertainties in the value of the LEC, $\tilde{e}_V^R(m_N^2)$, see text.

**Figure 2:** Radiative correction, $\frac{\alpha}{2\pi}\mathcal{G}_1^{rad}$, Eq. (8), and recoil correction, $m_N^{-1}\mathcal{G}_1^{recoil}$, Eq. (11), are plotted as functions of the incoming anti-neutrino energy $E_\nu$ for $\bar{\nu}_e + p \rightarrow e^+ + n$. The two RC curves indicate the "error band", see text. The vertical dotted line is the threshold ($E_\nu \simeq 1.8$ MeV).

$\mathcal{G}_1^{recoil}$ is enhanced by the fact that there is near cancellation among the various terms contributing to $\mathcal{K}_1^{recoil}$, whereas they add for $\mathcal{G}_1^{recoil}$, as is evident by comparing Eqs.(10) and (11). We also plot the "total recoil" correction, which is the combined effect of $m_N^{-1}$ terms in the lagrangian, Eq.(4), and $m_N^{-1}$ correction in phase space. We find after angle-integration

$$\frac{d\Gamma_\beta}{dE} = \frac{d\Gamma_\beta^{(0)}}{dE}\left\{1 + \frac{\alpha}{2\pi}\mathcal{K}_1^{rad}(\beta) + \frac{1}{m_N}\mathcal{K}_1^{\text{total}-\text{recoil}}(\beta)\right\}, \quad (12)$$

where $d\Gamma_\beta^{(0)}/dE = (G_F V_{ud})^2 \left[F(Z=1,E)E^2\beta(E^{max}-E)^2/(2\pi^3)\right](1+3g_A^2)$ is the LO result and

$$\mathcal{K}_1^{\text{total}-\text{recoil}}(\beta) = \mathcal{K}_1^{recoil}(\beta) + 3E - E^{max} - \left(\frac{1-g_A^2}{1+3g_A^2}\right)\beta^2 E + \mathcal{O}(m_N^{-1}). \quad (13)$$

Meanwhile, Eq.(5) (after integrated over the angle) can be rewritten as

$$\sigma_{inv-\beta} = \sigma_{inv-\beta}^{(0)}\left\{1 + \frac{\alpha}{2\pi}\mathcal{G}_1^{rad}(\beta) + \frac{1}{m_N}\mathcal{G}_1^{\text{total}-\text{recoil}}(\beta)\right\}, \quad (14)$$

where $\sigma_{inv-\beta}^{(0)} = (G_F V_{ud})^2(\tilde{E}^2\tilde{\beta}/\pi)(1+3g_A^2)$ is the LO result, and

$$\mathcal{G}_1^{\text{total}-\text{recoil}}(\beta) = \mathcal{G}_1^{recoil}(\beta) - \tilde{E} - \left(\frac{1+\tilde{\beta}^2}{\tilde{\beta}^2}\right)\left(E_\nu + \frac{\Delta_N^2 - m_e^2}{2\tilde{E}}\right) + \left(\frac{1-g_A^2}{1+3g_A^2}\right)E_\nu + \mathcal{O}(m_N^{-1}), \quad (15)$$

where $\tilde{E} = E_\nu - (m_n - m_p) = E_\nu - \Delta_N$ is the "LO positron energy" and $\tilde{\beta}$ the corresponding velocity.[3]

---

[3] Here, $E$ and $\beta$ in Eq.(6) have been expanded in powers of $m_N^{-1}$, see Ref. [12]. Although the $m_N^{-1}$-expanded expression in Eq. (15) diverges at the threshold, $\sigma_{inv-\beta} = 0$ at the threshold when the correct phase space $f(E)$ is used.

We note that $|\mathcal{K}_1^{\text{total}-\text{recoil}}|$ is significantly smaller than $|\mathcal{G}_1^{\text{total}-\text{recoil}}|$. Fig.1 indicate that the recoil corrections are much smaller than the RCs for neutron $\beta$-decay. By contrast, $\mathcal{G}_1^{rad}$ and $\mathcal{G}_1^{\text{total}-\text{recoil}}$ are of comparable size and have opposite signs. Fig. 2 shows that the recoil corrections and the RCs have very distinct energy dependencies. Thus both corrections must be carefully considered in analyzing high-precision data used to deduce $\theta_{13}$.

To summarize, we have presented a brief discussion of numerical consequences of the radiative corrections (RCs) and the recoil corrections for neutron $\beta$-decay, and inverse $\beta$-decay, calculated up to next-to-leading order (NLO) in HB$\chi$PT. The numerical results reported here are obtained with the use of the value of the LEC, $\tilde{e}_V^R(\mu^2)$, that has been deduced from comparison of our HB$\chi$PT results with those obtained in the S-M approach. This hybrid nature of our analysis should eventually be replaced by a more rigorous determination of $\tilde{e}_V^R(\mu^2)$ through a direct fit to, e.g., the neutron $\beta$-decay data. It is expected, however, that the basic features of the RCs are already visible in the present hybrid treatment. The recoil corrections are not affected by the hybrid nature.

This work is supported in part by the NSF grants, PHYS-0758114 and PHY-1068305.

# References

[1] Y. Abe *et al*. (Double Chooz Collaboration), Phys. Rev. Lett., **108**, 131801 (2012); arXiv:1112.6353.

[2] F.P. An *et al*. (Daya Bay Collaboration), Phys. Rev. Lett., **108**, 171803(2012); arXiv:1203.1669;

[3] J.K. Ahn *et al*. (RENO Collaboration), Phys. Rev. Lett., **108**, 191802 (2012).

[4] P. Adamson *et al*. (MINOS Collaboration), Phys. Rev. D **82**, 051102 (2010); Phys. Rev. Lett. **106**, 181801 (2011); Phys. Rev. Lett. **107**, 021801 (2011).

[5] M. Fukugita and T. Kubota, Acta Phys. Polon. **B35**, 1687 (2004)[arXiv:hep-ph/0403149]; Erratum: arXiv:hep-ph/0403149

[6] P. Vogel, Phys. Rev. D29, 1918 (1984)

[7] S.A. Fayans, Yad. Fiz. **42**, 929 (1985) [Sov. J. Nucl. Phys. **42**, 590 (1985)].

[8] A. Sirlin, Nucl. Phys. B**71**, 29 (1974); Nucl. Phys. B**100**, 291 (1975); arXiv:hep-ph/0309187 (2003)

[9] W.J. Marciano and A. Sirlin, Phys. Rev. Lett. **56**, 22 (1986).

[10] P. Vogel and J.F. Beacom, Phys. Rev. D **60**, 053003 (1999).

[11] S. Ando *et al*., Phys. Lett. B **595**, 250 (2004).

[12] U. Raha, F. Myhrer, K. Kubodera, Phys. Rev. C**85**, 045502 (2012); Erratum: Phys. Rev. C**86**, 039903 (2012).

[13] A. Sirlin, Phys. Rev. D **84**, 014021 (2011); arXiv:1105.2842 [hep-ph].

[14] V. Bernard, N. Kaiser and U.-G. Meißner, Int. J. Mod. Phys. **E4**, 193 (1995).

[15] S. Scherer, Adv. Nucl. Phys., **27**, 2001 (2003).

[16] V. Bernard, Prog. Part. Nucl. Phys. **60**, 82 (2008).

[17] W.J. Marciano and A. Sirlin, Phys. Rev. Lett. **96**, 032002 (2006).